\documentclass[twocolumn,english,prb]{revtex4}
\usepackage[T1]{fontenc}
\usepackage[latin9]{inputenc}
\usepackage{units}
\usepackage{amsmath}
\usepackage{graphicx}
\usepackage{amssymb}

\makeatletter
\@ifundefined{textcolor}{}
{%
 \definecolor{BLACK}{gray}{0}
 \definecolor{WHITE}{gray}{1}
 \definecolor{RED}{rgb}{1,0,0}
 \definecolor{GREEN}{rgb}{0,1,0}
 \definecolor{BLUE}{rgb}{0,0,1}
 \definecolor{CYAN}{cmyk}{1,0,0,0}
 \definecolor{MAGENTA}{cmyk}{0,1,0,0}
 \definecolor{YELLOW}{cmyk}{0,0,1,0}
 }

\makeatother

\makeatother

\usepackage{babel}

\makeatother

\usepackage{babel}

\makeatother

\usepackage{babel}

\makeatother

\usepackage{babel}

\makeatother

\usepackage{babel}

\begin{document}

\title{Quasiparticle relaxation dynamics in spin-density-wave and superconducting
SmFeAsO$_{1-x}$F$_{x}$ single crystals}

\author{T. Mertelj$^{1}$, P. Kusar$^{1}$, V.V. Kabanov$^{1}$, L. Stojchevska$^{1}$,
N.D. Zhigadlo$^{2}$, S. Katrych$^{2}$, Z. Bukowski$^{2}$, J. Karpinski$^{2}$,
S. Weyeneth$^{3}$ and D. Mihailovic$^{1}$}

\affiliation{$^{1}$Complex Matter Dept., Jozef Stefan Institute, Jamova 39, Ljubljana,
SI-1000, Ljubljana, Slovenia }

\affiliation{$^{2}$Laboratory for Solid State Physics, ETH Zürich, 8093 Zürich,
Switzerland}

\affiliation{$^{3}$Physik-Institut der Universität Zürich, 8057 Zürich, Switzerland}

\date{\today}
\begin{abstract}
We investigate the quasiparticle relaxation and low-energy electronic
structure in undoped SmFeAsO and near-optimally doped SmFeAsO$_{0.8}$F$_{0.2}$
single crystals - exhibiting spin-density wave (SDW) ordering and
superconductivity respectively - using pump-probe femtosecond spectroscopy.
In the undoped single crystals a single relaxation process is observed,
showing a remarkable critical slowing down of the QP relaxation dynamics
at the SDW transition temperature $T_{SDW}\simeq125\mbox{K}$. In
the superconducting (SC) crystals multiple relaxation processes are
present, with distinct SC state quasiparticle recombination dynamics
exhibiting a BCS-like $T$-dependent superconducting gap, and a pseudogap
(PG)-like feature with an onset above 180K indicating the existence
of a pseudogap of magnitude $2\Delta_{\mathrm{PG}}\simeq120$ meV
above $T_{\mathrm{c}}$. From the pump-photon energy dependence we
conclude that the SC state and PG relaxation channels are independent,
implying the presence of two separate electronic subsystems. We discuss
the data in terms of spatial inhomogeneity and multi-band scenarios,
finding that the latter is more consistent with the present data.
\end{abstract}
\maketitle
The discovery of high-temperature superconductivity in iron-based
pnictides (IP) \cite{KamiharaKamihara2006,kamiharaWatanabe2008,RenChe2008}
has attracted a great deal of attention recently. Contrary to the
cuprate superconductors, where a single band with a high degree of
correlations is believed to be sufficient starting point for the description
of the electronic properties, there is a clear theoretical\cite{singhDu2008}
and experimental\cite{coldeaFletcher2008} evidence that in IP several
bands cross the Fermi energy ($\epsilon_{\mathrm{F}}$).%
\footnote{For a recent review see ref. \onlinecite{IshidaNakai2009}.%
}The implications of the presence of several bands at $\epsilon_{\mathrm{F}}$
in IP are still under intense investigation. In the undoped state
two SDW gaps were detected by optical spectroscopy\cite{HuDong2008}
in 122 compounds (AFe$_{2}$As$_{2}$ A=Ba,Sr) presumably originating
from different bands crossing $\epsilon_{\mathrm{F}}$. In LaFeAsO
(La-1111) a similar optical conductivity suppression was observed\cite{ChenYuan2009},
but no analysis in terms of SDW gaps was performed. In the superconducting
state multiple superconducting gaps were detected\cite{DingRichard2008,KawasakiShimada2008,DagheroTortello2009,MartinTillman2009}
corresponding to different bands crossing $\epsilon_{\mathrm{F}}$.
In addition to superconducting gaps also the presence of a pseudogap
was reported by NMR\cite{AhilanNing2009,NakaiKitagawa2009} and point-contact
Andreev spectroscopy\cite{GonnelliDaghero2009} in La-1111. 

Time resolved spectroscopy has been very instrumental in elucidating
the nature of the electronic excitations in superconductors, particularly
cuprates, by virtue of the fact that different components in the low-energy
excitation spectrum could be distinguished by their different lifetimes
\cite{StevensSmith1997,KabanovDemsar99,DemsarPodobnik1999,KaindlWoerner2000,DvorsekKabanov2002,SegreGedik2002,SchneiderDemsar2002,DemsarAveritt2003,GedikOrenstein2003,GedikBlake2004,KusarDemsar2005,KaindlCarnahan2005,KabanovDemsar2005,BianchiChen2005,SchneiderOnellion2005}.
Moreover, the relaxation kinetics can give us valuable information
on the electronic density of states \cite{KabanovDemsar99} and electron-phonon
coupling\cite{KabanovAlexandrov2008}. Extensive and systematic experiments
on cuprates have also given information on the behavior of the pseudogap
for charge excitations, complementing the information obtained on
spin excitations from NMR and other spectroscopies \cite{DemsarPodobnik1999,DvorsekKabanov2002,KusarDemsar2005}. 

In this work we present a time-resolved femtosecond spectroscopy study
of undoped and near-optimally doped SmFeAsO$_{1-x}$F$_{x}$ single
crystals with $x=0$ and $x\simeq0.2$ with the aim of elucidating
the low energy electronic structure, investigating possible multi-component
response as a sign of phase separation and to obtain detailed information
about the quasiparticle (QP) dynamics in the normal, SDW and superconducting
states.

\section{Experimental}

Optical experiments were performed using the standard pump-probe technique,
with 50 fs optical pulses from a 250-kHz Ti:Al$_{2}$O$_{3}$ regenerative
amplifier seeded with an Ti:Al$_{2}$O$_{3}$ oscillator. We used
the pump photons with either doubled ($\hbar\omega_{\mathrm{P}}=3.1$
eV) or fundamental ($\hbar\omega_{\mathrm{P}}=1.55$ eV) photon energy
and the probe photons with 1.55 eV photon energy. The pump and probe
polarizations were perpendicular to each other and oriented with respect
to the the crystals to obtain the maximum amplitude of the response
at low temperatures. The pump and probe beam diameters were determined
by measuring the transmittance of calibrated pinholes mounted at the
sample place\cite{KusarKabanov2008}.

The crystals were flux grown at high pressure at ETH in Zurich\cite{ZhigadloKatrych2008}
and were approximately $\sim$120 x $\sim$80 $\mu$m in size. For
optical measurements the crystals were glued on a sapphire window
mounted in an optical liquid-He flow cryostat.

\subsection{Undoped, spin-density-wave ordered SmFeAsO}

\begin{figure}[tbh]
\begin{centering}
\includegraphics[width=0.3\textwidth]{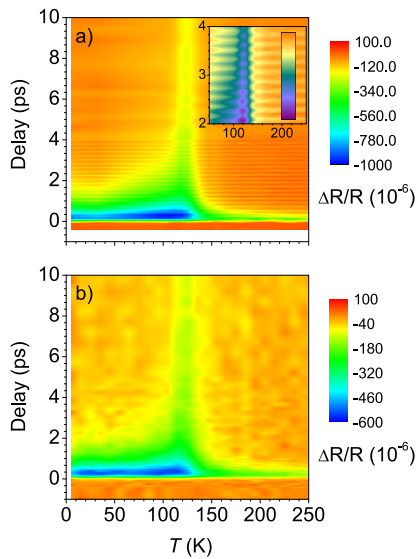} 
\par\end{centering}

\caption{$\Delta R/R$ transients as a function of temperature at 1.55 eV pump-photon
energy and 18 $\mu$J/cm$^{2}$ (a) and 3.1 eV pump-photon energy
and 15 $\mu$J/cm$^{2}$ (b) in undoped SmFeAsO. In (a) the coherent
phonon, shown expanded in the insert, is artificially smeared beyond
4 ps delay due to a decreased time resolution of scans.}

\label{fig:fig-400-SDW} 
\end{figure}

\begin{figure}[tbh]
\begin{centering}
\includegraphics[width=0.4\textwidth]{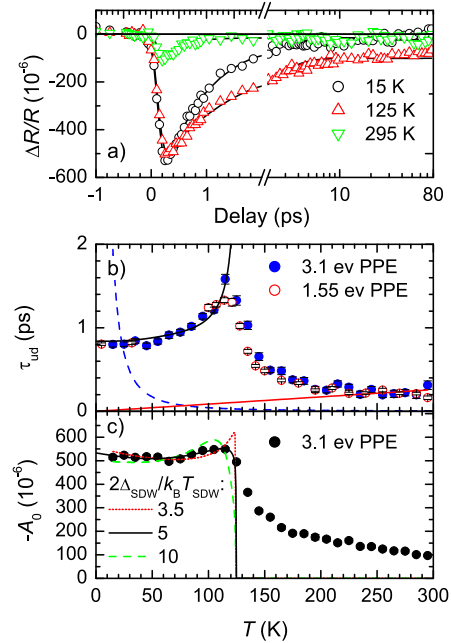} 
\par\end{centering}

\caption{$\Delta R/R$ transients at representative temperatures in undoped
SmFeAsO with single-exponential decay fits (a). The relaxation time
at two pump photon energies b) and amplitude (c) as functions of temperature
at $\mathcal{F}=15$ $\mu$J/cm$^{2}$. The red solid line in (b)
is fit of equation (\ref{eq:TauPoorHigh}) to $\tau_{\mathrm{ud}}$
above 230K. The blue dashed line in (b) is equation (\ref{eq:TauPoorLow})
with $\lambda=0.2$ and $\Theta_{\mathrm{D}}=175$K. Black thin solid
line in (a) represents the fit of equation (28) from \cite{KabanovDemsar99}discussed
in detail in text. Thin lines in (b) represent the fits of equation
(6) from \cite{KabanovDemsar99} with different magnitudes of the
gap. }

\label{fig:fig-AvsT-SDW} 
\end{figure}

In Fig. \ref{fig:fig-400-SDW} we plot temperature dependence of $\Delta R/R$
transients in undoped SmFeAsO. The only discernible difference of
the response at different pump-photon energies is the presence of
a coherent phonon oscillation with the frequency 5.1 THz (170 cm$^{-1}$)
at 295K, at $\hbar\omega_{\mathrm{P}}=1.55$ eV, which is absent at
$\hbar\omega_{\mathrm{P}}=3.1$ eV, consistent with Raman data\cite{HadjievIliev2008}.
Apart of the coherent phonon oscillation the transients consist from
a negative-amplitude single-exponential relaxation with a temperature
independent rise time of $\sim180$ fs (see Fig. \ref{fig:fig-AvsT-SDW}
(a)). Around $\sim125$K an additional long lived response appears
with decay time beyond our measurement delay range. The amplitude
of the transients, $A_{0}$, linearly increases with decreasing temperature
down to $\sim170$ K (see Fig. \ref{fig:fig-AvsT-SDW} (c)) with the
relaxation time, $\tau_{\mathrm{ud}}$, of $\sim220$ fs time being
virtually constant above 200K. Below 200 K $\tau_{\mathrm{ud}}$ starts
to increase while the amplitude starts to depart from the linear dependence
only below $\sim170$ K rapidly increasing below $\sim140$ K, and
achieving a maximum at $115$ K, upon entering the SDW state. With
further decrease of the temperature the amplitude slightly drops at
first and then remains constant below 50 K. Simultaneously with the
maximum of the amplitude $\tau_{\mathrm{ud}}$ shows a remarkable
divergent-like peak at $\sim120$ K and then drops to a temperature
independent value of 0.8 ps below 50 K.

\begin{figure}[tbh]
\begin{centering}
\includegraphics[width=0.35\textwidth]{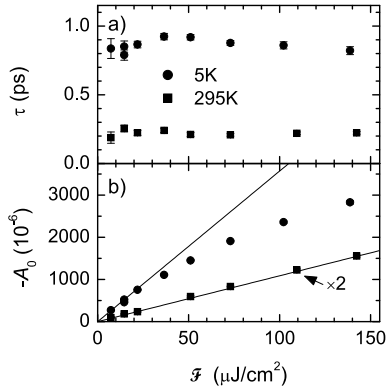} 
\par\end{centering}

\caption{Fluence dependence of the $\Delta R/R$ transient amplitude and relaxation
time in undoped SmAsFeO at two different temperatures.}

\label{fig:fig-SDW-AvsF-400} 
\end{figure}

The rise and decay times are virtually independent of the fluence,
$\mathcal{F}$, at all temperatures (see Fig. \ref{fig:fig-SDW-AvsF-400})
while the amplitude increases linearly with $\mathcal{F}$ at 295K
and shows a weak saturation above $\mathcal{F}=25$ $\mu$J/cm$^{2}$
at 5K.


\subsection{Superconducting SmFeAsO$_{0.8}$F$_{0.2}$}

\begin{figure}[tbh]
\begin{centering}
\includegraphics[width=0.3\textwidth]{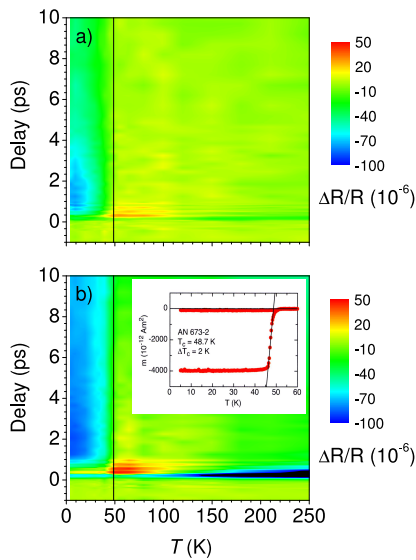} 
\par\end{centering}

\caption{$\Delta R/R$ transients as a function of temperature at 1.55-eV pump-photon
energy (a) and 3.1-eV pump-photon energy (b) in superconducting SmFeAsO$_{0.8}$F$_{0.2}$.
The pump fluence was 17 $\mu$J/cm$^{2}$ at 3.1 eV and 15 $\mu$J/cm$^{2}$
at 1.55 eV. Below $T_{\mathrm{c}}$ the response of the superconducting
state is clearly seen. The temperature dependence of the magnetization
is shown in the inset.}

\label{fig:fig-400-800-2D} 
\end{figure}

The temperature dependence of the $\Delta R/R$-transients in the
near optimally doped sample is shown in Fig. \ref{fig:fig-400-800-2D}.
Contrary to the undoped case the transients show complex time and
temperature dependencies. Independent of the $\hbar\omega_{\mathrm{P}}$
one can clearly resolve three temperature regions with different characteristic
behaviors.

\begin{figure}[tbh]
\begin{centering}
\includegraphics[bb=40bp 28bp 486bp 827bp,width=0.3\textwidth]{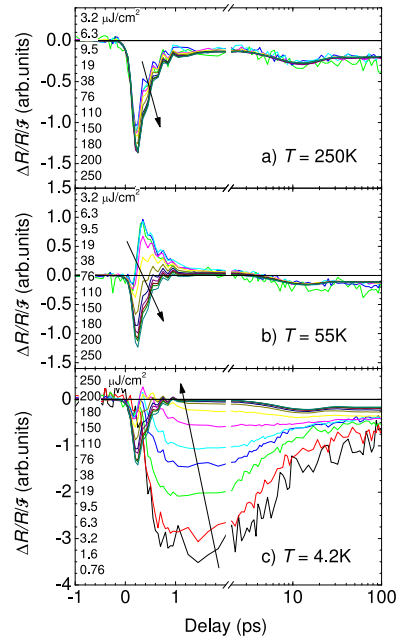} 
\par\end{centering}

\caption{Normalized $\Delta R/R$ transients at selected temperatures at 3.1-eV
pump photon energy as a function of $\mathcal{F}$ in superconducting
SmFeAsO$_{0.8}$F$_{0.2}$. Above $\sim150$ $\mu$J/cm$^{2}$ the
traces start to overlap indicating a linear $\mathcal{F}$-dependence.
The arrows indicate the direction of increasing $\mathcal{F}$.}

\label{fig:fig-F-dep-400} 
\end{figure}

\begin{figure}[tbh]
\begin{centering}
\includegraphics[width=0.3\textwidth]{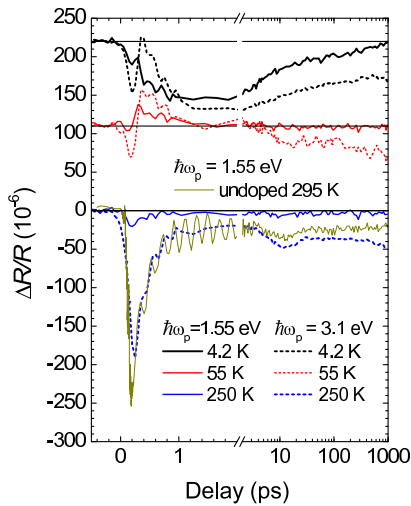} 
\par\end{centering}

\caption{$\Delta R/R$ transients at different pump-photon energies and selected
temperatures in superconducting SmFeAsO$_{0.8}$F$_{0.2}$. For comparison
the 295 K transient from undoped SmFeAsO is also shown.}

\label{fig:fig-400-800-T-selected} 
\end{figure}

(i) At the high temperatures a negative transient is observed with
initial 0.25-ps decay followed by a slower response consisting from
a weak peak at 12 ps (see Fig. \ref{fig:fig-F-dep-400} (a)) at $\hbar\omega_{\mathrm{P}}=3.1$
eV. The transients linearly scale with increasing fluence except in
the region of the initial 0.25-ps decay where a weak $\mathcal{F}$-dependence
is observed at low $\mathcal{F}$. At $\hbar\omega_{\mathrm{P}}=3.1$
eV the transients have the same shape and similar amplitude as in
undoped SmFeAsO without the coherent phonon. At $\hbar\omega_{\mathrm{P}}=1.55$
eV the negative hight-$T$ transients are much weaker than at $\hbar\omega_{\mathrm{P}}=3.1$
eV so only the initial 0.25-ps decay is resolved from the noise (see
Fig. \ref{fig:fig-400-800-T-selected}). In addition a coherent phonon
is observed with a softer frequency than in undoped SmFeAsO of 4.6
THz (153 cm$^{-1})$ having similar amplitude at both $\hbar\omega_{\mathrm{P}}$.

(ii) At the intermediate temperatures above $T_{c}$ and low $\mathcal{F}$
the transients are positive on the sub-ps timescale cross zero around
4 ps with slow dynamics similar to the high-temperature one. At high
$\mathcal{F}$ the positive part of the transients vanishes and the
transients become qualitatively identical to those at higher temperatures
(see Fig. \ref{fig:fig-F-dep-400}). The only remaining difference
is a delay-independent positive vertical shift of the the intermediate
$T$ scans with respect to those measured at 250 K and an increased
coherent phonon frequency of 5.1 THz (170 cm$^{1}$). The $\hbar\omega_{\mathrm{P}}=1.55$-eV
transients are, as at higher temperatures, similar to the $\hbar\omega_{\mathrm{P}}=3.1$-eV
transients but weaker.

(iii) Below $T_{\mathrm{c}}$ an additional negative component appears
with a rise time of 0.2-0.6 ps, depending on the $\hbar\omega_{\mathrm{P}}$
and pump fluence, and decay time of $\sim5$ ps. The component has
a similar amplitude at both $\hbar\omega_{\mathrm{P}}$. At high $\mathcal{F}$
the additional negative component becomes undetectable due to a saturation
and the transients become virtually identical to those measured at
55 K including the coherent phonon response.

\section{Discussion}

\subsection{Undoped SmFeAsO}

Upon cooling undoped LaFeAsO undergoes a sequence of a structural
transition from a tetragonal to orthorhombic symmetry, at $T_{\mathsf{s}}=156$K,
and a magnetic SDW transition at $T_{\mathrm{SDW}}$$=138$K.\cite{KlaussLuetkens2008}
In SmFeAsO $T_{\mathrm{SDW}}$$\sim135$K\cite{DrewNiedermayer2009}
while the structural transition was reported at lower temperature
$T_{\mathsf{s}}=130$K.\cite{MargadonnaTakabayashi2009} So far due
to possible different oxygen deficiencies in the two experiments\cite{MargadonnaTakabayashi2009,DrewNiedermayer2009}
and strong doping dependence of both $T_{\mathrm{SDW}}$ and $T_{\mathsf{s}}$
it was not possible to reliably distinguish between $T_{\mathrm{SDW}}$
and $T_{\mathsf{s}}$ in SmFeAsO. Our data show a marked critical
slowing down at $125$K (see Fig. \ref{fig:fig-400-SDW}) in the form
of the long-lived relaxation tail, while the initial picosecond exponential
decay time shows a maximum at $\sim115$K. Since the long-lived relaxation
tail affects the quality of the single-exponential picosecond fit
(see Fig. \ref{fig:fig-AvsT-SDW}(a)) we can not reliably identify
115 K as a separate transition temperature. We are therefore unable
to differentiate between the structural and spin transitions so we
will only refer to a single transition temperature $T_{\mathrm{SDW}}\approx T\mathrm{_{s}}\approx125$
K in the rest of the paper.

From our data $T_{\mathrm{SDW}}$ is lower than reported in literature\cite{DrewNiedermayer2009,SannaRenzi2009}.
The apparent lower $T_{\mathrm{SDW}}$ can originate in an elevated
temperature of the excited volume with respect to the cryostat temperature
due to the laser heating. The $\sim10$-K shift of $T_{\mathrm{SDW}}$
is larger than the shift of $T\mathrm{_{c}}$ of a few (2-3) K observed
in the superconducting crystals under similar excitation conditions.
Due to a variable thermal coupling to the sapphire substrate for such
small crystals a small decrease of $T_{\mathrm{SDW}}$ originating
from the oxygen deficiency can not be distinguished from the laser
heating in our samples.

Above $T_{\mathrm{SDW}}$ undoped pnictides are bad metals with resistivities
in the m$\Omega$cm range\cite{LuoWang2008,ChenYuan2009} and plasma
frequency in an $\sim1$ eV range\cite{ChenYuan2009,HuDong2008}.
The $\Delta R/R$ transients in this temperature range are therefore
attributed to the relaxation of electrons in the states near $\epsilon_{\mathrm{F}}$
and can be analyzed by means of the recent theoretical results\cite{KabanovAlexandrov2008}
on electron relaxation in metals. The $\mathcal{F}$-independent relaxation
time warrants use of the low excitation expansion\cite{KabanovAlexandrov2008},
where in the high temperature limit the relaxation time is proportional
to the temperature,\cite{GadermaierAlexandrov2009}\begin{equation}
\tau=\frac{2\pi k_{\mathrm{B}}T}{3\hbar\lambda\langle\omega^{2}\rangle}.\label{eq:TauPoorHigh}\end{equation}
Here $\lambda\langle\omega^{2}\rangle$ is the second moment of the
Eliashberg function\cite{KabanovAlexandrov2008}, $\lambda$ the electron-phonon
coupling constant and $k_{\mathrm{B}}$ the Boltzman constant. In
the low temperature limit the relaxation time is predicted to diverge
at low $T$:\cite{KabanovAlexandrov2008}\begin{equation}
\tau=\frac{2\hbar\Theta\mathrm{_{D}}}{\pi^{3}\lambda k_{\mathrm{B}}T^{2}},\label{eq:TauPoorLow}\end{equation}
where $\Theta_{\mathrm{D}}$ is the Debye temperature. From the fit
of equation (\ref{eq:TauPoorHigh}) to the relaxation time above 230
K (see Fig. \ref{fig:fig-AvsT-SDW}) we obtain $\lambda\langle\omega^{2}\rangle$=$135\pm10$
meV$^{2}$. If we estimate $\langle\omega^{2}\rangle\approx25^{2}$
meV$^{2}$ from inelastic neutron data\cite{OsbornRosenkranz2009}
we obtain $\lambda\approx0.2$ indicating a rather weak electron phonon
coupling, which can not explain high $T\mathrm{_{c}}$ in the doped
compound within a single band BCS model. However, owing to the multiband
nature of iron-pnictides it is possible, that due to optical selection
rules some bands with possible higher couplings are not detected by
$\Delta R/R$ transients. To check the consistency of the resulting
value of $\lambda$ we plot in Fig. \ref{fig:fig-AvsT-SDW} (b) also
the low-$T$ result (\ref{eq:TauPoorLow}) indicating validity of
the high-$T$ approximation (\ref{eq:TauPoorHigh}) above 230K.

Below $T_{\mathrm{SDW}}$ a gap opens at the Fermi surface introducing
a bottleneck in the relaxation. The relaxation across a temperature
dependent gap was analyzed by Kabanov \textit{et al.}.\cite{KabanovDemsar99}
We use equation (6) from Kabanov \textit{et al.}\cite{KabanovDemsar99},
which describes the photo-excited change in quasiparticle density
in the presence of a temperature dependent gap, to fit the amplitude
below $T_{\mathrm{SDW}}=125$ K. Using a single SDW gap energy with
the BCS temperature dependence and $2\Delta_{\mathrm{SDW}}/k_{\mathrm{B}}T_{\mathrm{SDW}}\simeq5$
results in a rather good fit to the amplitude temperature dependence
(see Fig. \ref{fig:fig-AvsT-SDW} (c)). Equation (28) for the relaxation
time from Kabanov \textit{et al.}\cite{KabanovDemsar99} with the
same $\Delta_{\mathrm{SDW}}\left(T\right)$ describes well also the
temperature dependence of the relaxation time (see Fig. \ref{fig:fig-AvsT-SDW}
(c)). However, equation (28) from Kabanov \textit{et al.}\cite{KabanovDemsar99}
also predicts a fast decrease of the relaxation time with $\mathcal{F}$,
which is not observed in our data. The reason for this might originate
in the fact that the SDW state is not fully gaped and the energy relaxation
is not limited by the anharmonic energy transfer from the high frequency
to the low frequency phonons as assumed in the derivation.\cite{KabanovDemsar99}

\subsection{Decomposition of the $\Delta R/R$ transients in superconducting
SmFeAsO$_{0.8}$F$_{0.2}$ into components.}

\begin{figure}[tbh]
\begin{centering}
\includegraphics[width=0.25\textwidth,angle=-90]{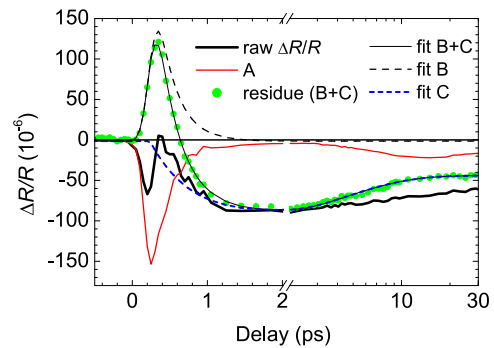} 
\par\end{centering}

\caption{Decomposition of $\Delta R/R$ transients into different components
in superconducting SmFeAsO$_{0.8}$F$_{0.2}$.}

\label{fig:fig-subtraction-4K2-400} 
\end{figure}

While the transients in the undoped sample show a simple single-exponential
relaxation, which is sensitive to the phase transition to the orthorhombic
SDW state, the transients in the doped sample show a clear muticomponent
relaxation. To separate contributions from different components we
use\cite{MerteljKabanov2009prl,MerteljKabanov2009jsnm} the fluence
dependence of the reflectivity transients. The data can be consistently
described by three distinct components A, B and C, which are tightly
connected with three observed temperature regions. 

The temperature-independent linear scaling of the transients with
$\mathcal{F}$ above $\sim150$ $\mu$J/cm$^{2}$ suggests the decomposition
of the raw $\Delta R/R$ into component A which scales linearly with
$\mathcal{F}$ and a \emph{residue} which saturates at finite $\mathcal{F}$
(see Fig. \ref{fig:fig-subtraction-4K2-400}). Component A dominates
at high temperatures and has to originate in at least three distinct
relaxation processes due to the relatively complex time evolution.
(i) The slower dynamics, which is virtually the same as in undoped
SmFeAsO at high temperatures (see Fig. \ref{fig:fig-400-800-T-selected}),
could be attributed to the band renormalization due to the lattice
expansion. (ii) The sub-ps decay, also having a similar decay time
as in undoped SmFeAsO at high temperatures, will be discussed in more
detail below. (iii) The oscillatory part of component A is attributed
to a coherent phonon oscillation which appears softer as in undoped
SmFeAsO. Except for the shift of the coherent phonon frequency all
show only a minor $T$-dependence.

\begin{figure}[tbh]
\begin{centering}
\includegraphics[bb=14bp 2cm 582bp 26cm,clip,width=0.35\textwidth]{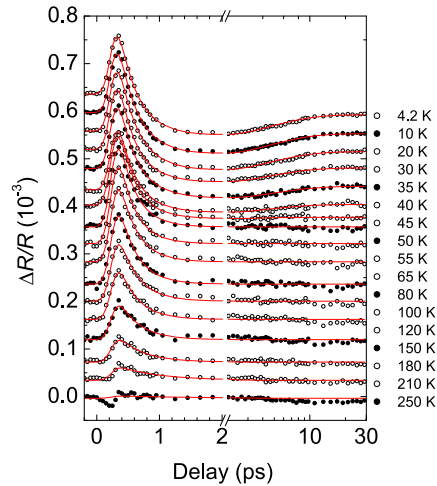} 
\par\end{centering}

\caption{Temperature dependence of $\Delta R/R$ transients with component
A subtracted. Thin lines are single exponential decay fits (component
B) above $T_{\mathrm{c}}$ and multi-exponential decay fits (a sum
of component B and C) below $T_{\mathrm{c}}$. }

\label{fig-17uJ-cm2-diff-fit} 
\end{figure}

The \emph{residue} shows an unipolar single-exponential decay (see
Fig. \ref{fig-17uJ-cm2-diff-fit}) above $T_{\mathrm{c}}$, which
we name component B. Component B dominates in the raw transients in
the intermediate temperature range, above $T_{\mathrm{c}}$. Below
$T_{\mathrm{c}}$ the \emph{residue} changes to a bipolar multi-exponential
decay, evidently due to the appearance of an additional relaxation
process associated with the superconducting state named component
C. Component C saturates at lower $\mathcal{F}\approx10$ $\mu$J/cm$^{2}$
than component B, which saturates above $\sim70$ $\mu$J/cm$^{2}$.
The decomposition to the three components is further supported by
comparison of the raw $\Delta R/R$ at different $\hbar\omega_{\mathrm{P}}$
shown in Fig. \ref{fig:fig-400-800-T-selected} where components A
and B show much smaller amplitudes at $\hbar\omega_{\mathrm{P}}=1.55$
eV in comparison to $\hbar\omega_{\mathrm{P}}=3.1$ eV, while the
amplitude of component C shows a negligible $\hbar\omega_{\mathrm{P}}$
dependence.

\begin{figure}[tbh]
\begin{centering}
\includegraphics[width=0.35\textwidth]{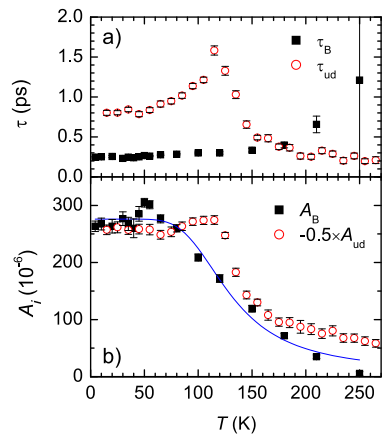} 
\par\end{centering}

\caption{Temperature dependence of the component-B relaxation time (a) and
amplitude (b) in superconducting SmAsFeO$_{0.8}$F$_{0.2}$ obtained
from the fits. The thin line is the fit for the case of a relaxation
over a $T$-independent gap\cite{KabanovDemsar99,MerteljKabanov2009prl}
with $2\Delta_{\mathrm{PG}}=120$meV. For comparison, the temperature
dependence of the relaxation time and the transient amplitude in undoped
SmAsFeO is also shown.}

\label{fig:fig-AvsT-B} 
\end{figure}

Above $T_{\mathrm{c}}$ we fit the \emph{residue} with a single exponential
decay (component B )\cite{MihailovicDemsar99} (see Fig. \ref{fig-17uJ-cm2-diff-fit}),\begin{equation}
\frac{\Delta R\mathrm{_{B}}}{R}=\frac{A_{\mathrm{B}}}{2}\mathrm{e}^{-\frac{t-t_{0}}{\tau_{\mathrm{B}}}}\operatorname{erfc}\left(\frac{\sigma^{2}-4(t-t_{0})\tau_{\mathrm{B}}}{2\sqrt{2}\sigma\tau_{\mathrm{B}}}\right),\label{eq:fitB}\end{equation}
 where $\sigma$ corresponds to the effective width of the excitation
pulse with a Gaussian temporal profile arriving at $t_{0}$ and $\tau_{\mathrm{B}}$
the exponential relaxation time. Below $T_{\mathrm{c}}$ additional
exponential decays representing component C are needed to fit the
\emph{residue}, \begin{equation}
\begin{split}\frac{\Delta R_{\mathrm{C}}}{R} & =A_{1\mathrm{C}}\left(\mathrm{e}^{-\frac{t-t_{0}}{\tau_{1\mathrm{C}}}}-\mathrm{e}^{-\frac{t-t_{0}}{\tau_{\mathrm{rC}}}}\right)\\
 & +A_{2\mathrm{C}}\left(\mathrm{e}^{-\frac{t-t_{0}}{\tau_{2\mathrm{C}}}}-\mathrm{e}^{-\frac{t-t_{0}}{\tau_{\mathrm{rC}}}}\right)\end{split}
\label{eq:fitC}\end{equation}
where $\tau_{\mathrm{rC}}$ represents the rise time and $\tau_{i\mathrm{C}}$
the decay times. The resulting fit parameters for component B are
shown in Fig. \ref{fig:fig-AvsT-B}. The decay time, $\tau_{1\mathrm{B}}\approx0.25$
ps at 4 K, slightly increases with increasing temperature below 200
K. Above 200 K, where artifacts due to subtraction of component A
start to be significant, $\tau_{1\mathrm{B}}$ steeply increases towards
1 ps. The amplitude of component B, $A_{\mathrm{B}}$, stays almost
constant up to $\sim70$ K and then drops monotonously.

\subsection{Superconducting response in SmFeAsO$_{0.8}$F$_{0.2}$}

\begin{figure}[tbh]
\begin{centering}
\includegraphics[angle=-90,width=0.45\textwidth]{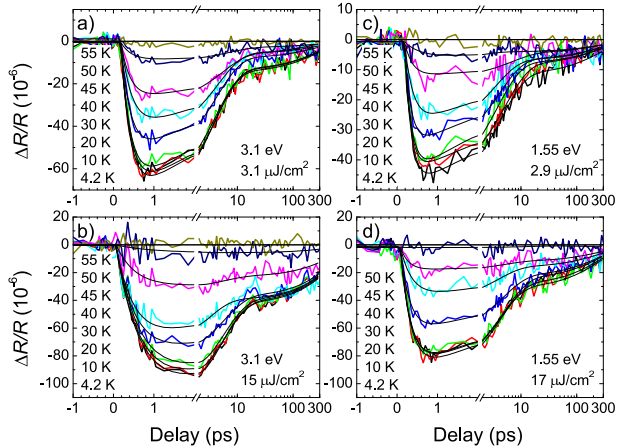} 
\par\end{centering}

\caption{Component C as a function of temperature at different pump photon
energies and fluences. Thin lines are fits discussed in text.}

\label{fig:fig-sc-response} 
\end{figure}

For easier separation of component C associated with the superconducting
response we use the fact that components A and B are temperature independent
in the superconducting state. We therefore extract component C by
subtracting the average of transients measured at 55 and 65 K from
transients measured below 55 K. The subtracted transients clearly
show a two-step decay with a finite rise time and are excellently
fit by equation (\ref{eq:fitC}) as shown in Fig. \ref{fig:fig-sc-response}.

In the spirit of the Rotwarf-Taylor model\cite{KabanovDemsar2005}
we associate the rise time with establishment of the thermal quasi-equilibrium
between the photo-excited quasiparticles and high frequency ($\hbar\omega>2\Delta_{\mathrm{SC}}$)
phonons. The shorter relaxation time is associated with establishment
of the local thermal equilibrium between all degrees of freedom, while
the longer relaxation time is due to energy escape out of the probed
volume. This is supported by increased relative amplitude of the long
decay with respect to the total amplitude at higher $\mathcal{F}$.

\begin{figure}[tbh]
\begin{centering}
\includegraphics[width=0.47\textwidth]{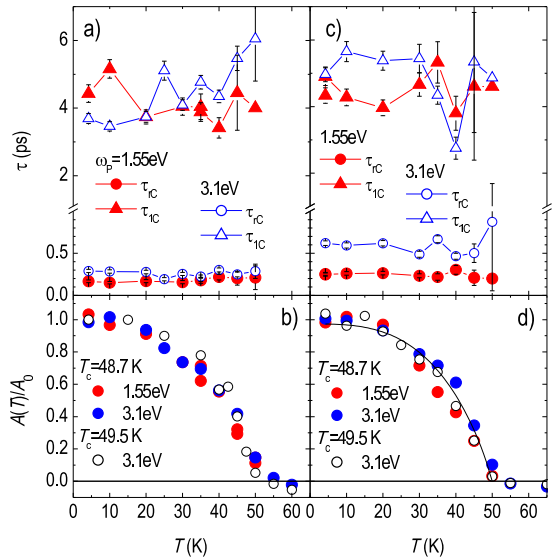} 
\par\end{centering}

\caption{Rise time and relaxation time (a), (c) and amplitude (b), (d) of component
C as functions of temperature at $\mathcal{F}=2.9$ $\mu$J/cm$^{2}$
and $3$ $\mu$J/cm$^{2}$ for $\hbar\omega_{\mathrm{P}}=1.55$ eV
and $3.1$ eV, respectively, (a), (b) and at $\mathcal{F}=17.4$ $\mu$J/cm$^{2}$
and $15$ $\mu$J/cm$^{2}$ for $\hbar\omega_{\mathrm{P}}=1.55$ eV
and 3.1 eV, respectively (c), (d). For comparison amplitudes in a
slightly higher $T\mathrm{_{c}}$ sample are shown in (b), (d). The
thin line in (c) is the fit of equation (\ref{eq:MattBard}) to the
data.}

\label{fig:fig-A-SC-T} 
\end{figure}

Neither the rise time nor the relaxation times (see Fig. \ref{fig:fig-A-SC-T})
show any temperature dependence within experimental error, while only
$\tau_{\mathrm{rC}}$ shows dependence on $\mathcal{F}$ and $\hbar\omega_{\mathrm{P}}$.
$\tau_{\mathrm{rC}}$ is always faster at $\hbar\omega_{\mathrm{P}}=1.55$
eV and shows much weaker $\mathcal{F}$-dependence than at $\hbar\omega_{\mathrm{P}}=3.1$
eV, where it increases from 0.3 ps at $\mathcal{F}=3$ $\mu$J/cm$^{2}$
to 0.6 ps at $\mathcal{F}=15$ $\mu$J/cm$^{2}$. While an increase
of $\tau_{\mathrm{rC}}$ with increasing $\hbar\omega_{\mathrm{P}}$
is expected due to the photo-excitation being farther away from $\epsilon_{\mathrm{F}}$
an increasing $\mathcal{F}$-dependence with increasing $\hbar\omega_{\mathrm{P}}$
is not completely understood.%

\begin{figure}[tbh]
\begin{centering}
\includegraphics[width=0.3\textwidth,angle=-90]{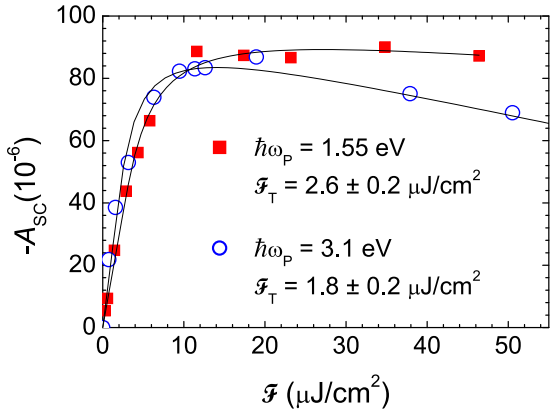} 
\par\end{centering}

\caption{The amplitude of the component C as functions of fluence at different
$\hbar\omega_{\mathrm{P}}$. Thin lines are fits discussed in text.}

\label{fig:fig-AvsF-SC} 
\end{figure}

In Fig. \ref{fig:fig-AvsF-SC} we plot $\mathcal{F}$-dependence of
the component C amplitude, $A_{\mathrm{SC}}$. Similarly as in the
cuprates\cite{KusarKabanov2008} the response saturates with increasing
$\mathcal{F}$ indicating a complete destruction of the superconducting
state in the excited volume. By taking into account the effects of
inhomogeneous excitation due to finite penetration depths and beam
diameters\cite{KusarKabanov2008} we determine the threshold external
fluence, $\mathcal{F}_{\mathrm{T}}$, at which the superconductivity
is destroyed in the most excited spot of the pump beam. From $\mathcal{F}_{\mathrm{T}}$
we calculate, using optical penetration depths, $\lambda_{\mathrm{op}}$,
and reflectivities, $R$, of LaAsFeO$_{1-x}$F$_{x}$,\cite{BorisKovaleva2009}
the energy density, $U\mathrm{_{p}}$, required to completely destroy
the superconducting state: $\nicefrac{U_{\mathrm{p}}}{k_{\mathrm{B}}}=\nicefrac{\mathcal{F}_{\mathrm{T}}\left(1-R\right)}{\lambda_{\mathrm{op}}k_{\mathrm{B}}}=2.2$
K/Fe ($U_{\mathrm{p}}=18$ J/mol) at $\hbar\omega_{\mathrm{P}}=3.1$
eV and $\nicefrac{U_{\mathrm{p}}}{k_{\mathrm{B}}}=1.5$ K/Fe ($U_{\mathrm{p}}=12$
J/mol) at $\hbar\omega_{\mathrm{P}}=1.55$ eV. The average value $\nicefrac{U_{\mathrm{p}}}{k_{\mathrm{B}}}=1.8$
K/Fe is slightly smaller than the values observed in La$_{1-x}$Sr$_{x}$CuO$_{4}$.\cite{KusarKabanov2008}
If we assume that the thermodynamic superconducting condensation energy,
$U_{\mathrm{c}}$, in SmFeAsO$_{0.8}$F$_{0.2}$ is similar to La$_{1-x}$Sr$_{x}$CuO$_{4}$
due to similar magnitudes of $T_{\mathrm{c}}$ we obtain $\nicefrac{U_{\mathrm{p}}}{U_{\mathrm{c}}}\gg1$
indicating that a significant amount of excitation energy is transferred
to the bath on a timescale of $\sim0.3$ ps. If all degrees of freedom
would absorb $U\mathrm{_{p}}$ the resulting temperature rise would
be 11K based on the published heat capacity, $c_{\mathrm{p}}$, data\cite{TropeanoMartinelli2008}.
Contrary to the cuprates $c_{\mathrm{p}}$ is dominated by Sm spins\cite{DingHe2008,TropeanoMartinelli2008}
below $\sim12$ K so it is not possible to determine whether the excess
$U\mathrm{_{p}}$ is absorbed in the Sm-spin or in the phonon subsystem.

We fit the temperature dependence of the reflectivity change upon
complete destruction of the superconducting state shown in Fig. \ref{fig:fig-A-SC-T}
(d) by the high-frequency limit of the Mattis-Bardeen formula,\cite{MattisBardeen1985}
\begin{equation}
\frac{\Delta R}{R}\propto\left(\frac{\Delta\left(T\right)}{\hbar\omega}\right)^{2}\log\left(\frac{3.3\hbar\omega}{\Delta\left(T\right)}\right),\label{eq:MattBard}\end{equation}
 where $\hbar\omega$ is the probe-photon energy and $\Delta\left(T\right)$
the superconducting gap. By using the BCS-gap temperature dependence
with $\nicefrac{2\Delta_{0}}{k_{\mathrm{B}}T_{\mathrm{c}}}=3.5$ we
obtain an excellent fit to the observed temperature dependence. Unfortunately
the shape of the temperature dependence (\ref{eq:MattBard}) is a
very weak function of $\nicefrac{2\Delta_{0}}{k_{\mathrm{B}}T_{\mathrm{c}}}$
and one can not reliably distinguish the contributions from different
gaps\cite{KarpinskiZhigadlo2009} and reliably determine $\nicefrac{2\Delta_{0}}{k_{\mathrm{B}}T_{\mathrm{c}}}$.%

\subsection{Normal state response in SmFeAsO$_{0.8}$F$_{0.2}$}

The temperature dependence of the component-B amplitude is consistent
with a bottleneck due to the relaxation over a $T$-independent gap\cite{KabanovDemsar99}
(see Fig. \ref{fig:fig-AvsT-B}) with a magnitude $2$$\Delta_{\mathrm{PG}}=120$
meV as noted previously.\cite{MerteljKabanov2009prl} The sub-ps part
of component A on the other hand is temperature independent suggesting
finite density of states at $\epsilon_{\mathrm{F}}$. This is consistent
with heat capacity measurements in polycrystalline SmFeAsO$_{1-x}$F$_{x}$
\cite{DingHe2008,TropeanoMartinelli2008} where a finite Sommerfeld
constant in the superconducting state suggests a finite density of
states at $\epsilon_{\mathrm{F}}$. Due to very similar pump-photon
energy dispersion of components A and B we believe that they originate
from the same electronic states which have a soft-gapped density of
states at $\epsilon_{\mathrm{F}}$. Saturation of component B amplitude
with increasing $\mathcal{F}$ indicates that the pseudogap can be
destroyed so it is not a simple band-structure effect. 

The pump-photon energy dispersion similar to that of components A
and B is not observed for component C. Electronic states involved
in the relaxation related to components A and B must therefore be
different than for component C. This confirms that the relaxation
below $T_{\mathrm{c}}$ does \emph{not} proceed via a cascade but
rather through distinct parallel channels as suggested previously.\cite{MerteljKabanov2009prl}
These channels correspond to two distinct electronic subsystems which
are weakly coupled on the sub-ps timescale. One exhibits the superconducting
gap(s) and the other a pseudogap.

A possible origin of the distinct electronic subsystems could be a
chemical phase separation of the doped F. However, the superconducting
transition is rather narrow (see Fig. \ref{fig:fig-400-800-2D}) so
the presence of weakly-superconducting fluorine-poor regions in which
SDW is suppressed giving rise to additional pseudo-gapped electronic
subsystem is also unlikely. More importantly undoped SmFeAsO shows
virtually no pump-photon energy dispersion which the superconducting
sample does, so a simple chemical phase separation to doped and undoped
regions is very unlikely despite similarity (see Fig. \ref{fig:fig-400-800-T-selected})
between component A and the undoped SmFeAsO room-temperature transients.
Moreover, there is no divergent signal at 125 K (or anywhere near
that temperature) which can be attributed to the presence of the undoped
phase in the superconducting sample. We therefore believe that both
electronic subsystems are intrinsic to the SC material

Apart from the chemical phase separation an intrinsic electronic phase
separation akin to that proposed for the cuprates\cite{Gorkov2001}
could be origin of the distinct electronic subsystems. The existence
of intrinsic electronic phase separation has been reported in 122
systems,\cite{ParkInosov2009,LaplaceBobroff2009} however at present
the issue is still rather controversial.

In the case of the spatially homogeneous electronic state different
electronic subsystems would correspond to different bands crossing
$\epsilon_{\mathrm{F}}$.$ $ This would imply that the inter-band
scattering between parts of the Fermi surface corresponding to different
electronic subsystems is negligible on a timescale of a few hundred
fs, since excitation photon at 1.5 eV only weakly excites components
A and B. Further, since both components exist in the superconducting
state even at low $\mathcal{F}$ the part of the Fermi surface corresponding
to components A and B has to remain ungapped or pseudo-gaped in the
superconducting state.

Another possibility for a weakly coupled electronic subsystem are
Sm crystal-field levels. The energy of the levels in SmFeAsO$_{1-x}$F$_{x}$
is in the range of 20-60 meV as determined indirectly from heat capacity
fits.\cite{CimberleFerdeghini2008,BakerGiblin2009} Involvement of
the crystal-field levels could explain the strong $\hbar\omega_{\mathrm{P}}$-dependence
of components A and B and decoupling from the other low lying electronic
states. However, the observation of a pseudogap by NMR in La-1111\cite{AhilanNing2009,NakaiKitagawa2009}
(which has no crystal field level structure at low energy), and the
weak $\hbar\omega_{\mathrm{P}}$-dependence in undoped SmFeAsO point
against such a scenario.

Finally, let us briefly compare our results in Sm-1111 to femtosecond
spectroscopy in (Ba,K)-122.\cite{ChiaTalbayev2008,TorchinskyChen2009}
There is a marked difference in the magnitude of the relaxation time
in the superconducting state, which increases with decreasing temperature
beyond 60 ps in optimally doped (Ba,K)-122\cite{ChiaTalbayev2008}
and remains $T$-independent in Sm-1111 at $\sim$5 ps. Similarly,
the excitation fluence dependence of the relaxation time, which is
absent in Sm-1111 is pronounced in (Ba,K)-122.\cite{TorchinskyChen2009}
This suggests different relaxation mechanisms in Sm-1111 and (Ba,K)-122.
While behavior in (Ba,K)-122 is consistent with the Rotwarf-Taylor
model\cite{RothwarfTaylor1967,KabanovDemsar2005} where the anharmonic
optical-phonon decay is determining the relaxation time in Sm-1111
the presence of the ungapped electronic subsystem seems to provide
a competing relaxation channel. However, only $\hbar\omega_{\mathrm{P}}=1.55$
eV was used in (Ba,K)Fe$_{2}$As$_{2}$ so ultrafast pump-probe spectroscopy
at different $\hbar\omega_{\mathrm{P}}$ in 122 systems and measurements
in LaFeAsO$_{1-x}$F$_{x}$ are needed for determination whether some
fundamental difference between 1111 and 122 systems is responsible
for the different relaxation time behavior and marked temperature
dependence above $T\mathrm{_{c}}$ in Sm-1111.

\section{Summary and conclusions}

In undoped SmFeAsO a single-exponential relaxation is observed. From
the high-$T$ relaxation time the second moment of the Eliashaberg
function is determined to be $\lambda\left\langle \omega^{2}\right\rangle =135\pm10$
meV$^{2}$. The coupling constant $\lambda\approx0.2$, estimated
from this value, is comparable to low-$T$$_{c}$ superconductors
and cannot explain the high superconducting $T_{c}$ s of these compounds
within a single band BCS model. 

Below $T_{\mathrm{SDW}}$ the temperature dependence of the relaxation
indicates appearance of a QP relaxation bottleneck due to opening
of a single charge gap at $T\mathrm{_{SDW}}$ with a BCS-like temperature
dependence and the amplitude of $2\Delta_{\mathrm{SDW}}/k_{\mathrm{B}}T_{\mathrm{SDW}}\simeq5$
at 4.2 K. A question whether this charge gap is a direct consequence
of the SDW formation or due to the structural transition unfortunately
cannot be answered from our data.

In superconducting SmFeAsO$_{0.8}$F$_{0.2}$ three distinct relaxation
components are observed. Components A and B are present in both the
superconducting and the normal state. The temperature dependence of
the amplitude of component B suggests the presence of a temperature
independent pseudogap with a magnitude $2$$\Delta_{\mathrm{PG}}\simeq120$
meV. The pseudogap is destroyed at a finite fluence indicating that
it is not a band-structure effect (such as a 120 meV gap at some arbitrary
point in the Brillouin zone). Component C is observed only in the
superconducting state and corresponds to the relaxation across a $T$-dependent
superconducting gap with a BCS temperature dependence. At high enough
pump fluence a complete destruction of the superconducting state is
observed with the critical optical excitation density $\nicefrac{U\mathrm{p}}{k_{\mathrm{B}}}\approx1.8$
K/Fe which is similar to the value observed in (La,Sr)CuO$_{4}$.

The multicomponent relaxation in SC samples strongly suggest the presence
of two relatively weakly coupled electronic subsystems, one exhibiting
the SC gap(s) and the other the pseudogap. From the temperature and
fluence dependence of photoinduced optical reflectivity transients
in undoped and near-optimally doped SmFeAsO$_{1-x}$F$_{x}$ single
crystals it is clear that the presence of two electronic subsystems
in the superconducting sample is not a result of a simple phase separation.
The fact that no relaxation component - such as appears in the SDW
phase - is seen in the SC phase appears to rule this out. The presence
of two electronic subsystems therefore originates either in an intrinsic
phase separation or more likely in the multiband nature of the superconducting
iron-pnictides.
\begin{acknowledgments}
This work has been supported by ARRS (GrantNo.P1-0040) and the Swiss
National Science Foundation NCCR MaNEP.
\end{acknowledgments}
\bibliography{biblio}

\end{document}